\documentclass[a4paper,11pt]{article}
\usepackage{pos}

\usepackage{tikz}

\title{A New Model for Jet Energy Loss in Heavy Ion Collisions}

\author*[a]{Alexander Lind}
\author[a,b]{Iurii Karpenko}
\author[c]{Martin Rohrmoser}
\author[a]{Joerg Aichelin}
\author[a]{Pol Gossiaux}
\author[a]{Klaus Werner}

\affiliation[a]{SUBATECH, Universit\'e de Nantes, IMT Atlantique, IN2P3/CNRS, \\
	4 rue Alfred Kastler, 44307 Nantes cedex 3, France}

\affiliation[b]{Czech Technical University in Prague, FNSPE, \\
	B\v{r}ehov\'a 7, Prague 11519, Czech Republic}

\affiliation[c]{Cracow University of Technology, Institute of Physics, \\
	Podchor\k{a}\.{z}ych 1, PL-30-084 Krak\'ow, Poland}

\emailAdd{alexander.lind@subatech.in2p3.fr}

\abstract{We present a new model for jet quenching from coherent radiation in a brick medium. The jet energy loss is simulated as a perturbative final-state vacuum parton shower followed by a medium-induced shower originating from elastic and radiative collisions with the medium constituents. Coherency is achieved by starting with trial gluons that act as field dressing of the initial jet parton. These are formed according to a Gunion-Bertsch seed. The QCD version of the LPM effect is attained by increasing the phase of the trial gluons through elastic scatterings with the medium. Above a phase threshold, the trial gluons will be realised and can produce coherent radiation themselves. The model has been implemented in a Monte Carlo code and has been validated by successfully reproducing the BDMPS-Z prediction for the energy spectrum. The realistic case with minimal assumptions are also produced and shown. In particular, we show the influence of various parameters on the energy spectrum and transverse momentum distribution, such as the in-medium quark masses, the energy transfer in the recoil process, and the phase accumulation criteria, especially for low and intermediate energy gluons. Future studies will allow for the interface with full simulations of the quark-gluon-plasma with hydrodynamic evolution, such as vHLLE, along with subsequent hadronisation of the jet partons in order to produce realistic distributions that can be directly compared to LHC and RHIC data.}

\FullConference{
 26-31 March 2023 \\
 Aschaffenburg, Germany\\}

\begin{document}

\maketitle


\section{Introduction}

In ultra-relativistic heavy ion collisions, a hot, dense, and deconfined state of quarks and gluons, called the \emph{quark-gluon-plasma} (QGP), can be created. Hard probes, such as jets, provides an excellent tool to study the QGP as the jet partons will interact with the QGP and carry information about the early stages of the collision, providing further understanding of the QCD phase diagram.

It is observed that the interactions of the jet with the QGP medium will result in modifications of the jet properties -- this is called \emph{jet energy loss} or \emph{jet quenching}. The calculation of this phenomenon is difficult owing to the hard, perturbative nature of the jet partons and the soft, non-perturbative, and collective behaviour of the QGP constituents. One has to rely on models for the description of medium modified jets. This motivates the development of Monte Carlo tools which can dynamically simulate the interactions between jets and the QGP medium.

We present a new model for jet energy loss in heavy ion collisions and its implementation in a Monte Carlo code called SUBA-Jet. We present first results, where we successfully reproduce the BDMPS-Z results for the energy spectrum and the transverse momentum distributions.


\section{Description of the algorithm}

This section provides a brief overview of the algorithm used in SUBA-Jet. A full description with technical details will be presented in \cite{paper}.

The algorithm is divided into two regimes -- a high virtuality and a low virtuality regime.


\subsection{High virtuality regime}

In the high virtuality regime, we emply a vacuum parton shower originally presented in \cite{martinthesis}. The evolution is according to the DGLAP equations from a high virtuality scale down to a low virtuality scale. The high virtuality scale can maximally be the $p_T$ of the hard scattering process producing the initial partons.

The simulation runs over time-steps $\Delta t$. During each time-step, each of the jet partons above the minimum virtuality $2 Q_0$ can split into two partons with lower virtualities according to a Sudakov form factor with a probability of $\Delta t \cdot Q^2 / E$. This ensures that the mean life-time of the partons between two successive splittings is $\tau = E / Q^2$.

Medium modifications of the jet partons in the high virtuality regime will be performed in a similar way to that of YaJEM \cite{Renk:2008pp}, where the virtuality will be increased each time-step according to a $\hat{q}$ parameterisation depending on parton momentum $p$ and medium temperature $T$. In practice, the virtuality increase is implemented using an energy increase,
\begin{equation}
	\frac{\mathrm{d}Q^2}{\mathrm{d}t} = \frac{\mathrm{d}E^2}{\mathrm{d}t} = \hat{q}(p,T) \,,
\end{equation}
such that the three-momentum of the partons are unchanged. The overall virtuality (energy) loss of the partons due to the parton splittings will be larger than the virtuality (energy) increase due to $\hat{q}$.

\begin{figure}[ht]
	\centering
	\begin{minipage}{0.49\textwidth}
		\centering
		\includegraphics[width=1.02\textwidth]{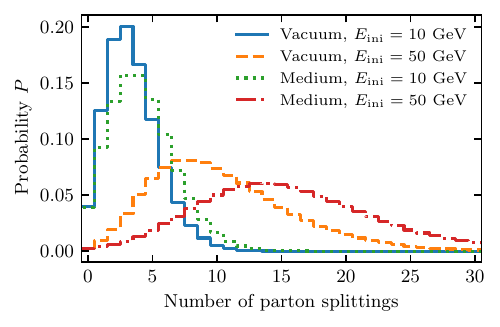}
		\caption{Probability that a parton vacuum cascade exhibits a given number of parton splittings for $E_{\text{ini}}=10$ and $50$~GeV, and with/without medium interactions.}
		\label{fig:ndist}
	\end{minipage}\hfill
	\begin{minipage}{0.49\textwidth}
		\centering
		\includegraphics[width=1.02\textwidth]{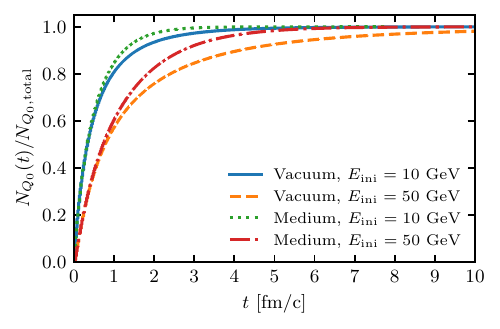}
		\caption{Distributions of the ratio of initial tagged quarks that have reached the minimum virtuality $2 Q_0$ as a function of time, for different initial energies and with/without medium modifications.}
		\label{fig:tq0ratio}
	\end{minipage}
\end{figure}

Figure~\ref{fig:ndist} shows the probability distributions of the number of parton splittings for different initial energies. The effect of medium modifications is also shown. The virtuality increase due to the medium modifications will result in more splittings and hence a larger overall energy loss.

Figure~\ref{fig:tq0ratio} shows a distribution of the ratio of initial tagged quarks that have reached the minimum virtuality as a function of time. The runs always start with a quark so the initial quark is tagged as the fermion in the $q \to qg$ splittings. It can be seen that after roughly $1$--$2$ fm$/$c, around $80\%$ of the initial tagged quarks has reached the minimum virtuality. We have taken here $Q_0 \sim \lambda_{\text{QCD}}$, but future studies will study the effect of larger values of $Q_0$.


\subsection{Low virtuality regime}

Once the jet partons have reached a minimum virtuality scale $Q_0^2$, they will enter the low virtuality regime of the algorithm where the now on-shell jet partons can undergo elastic and inelastic (radiative) interactions with the medium.

The elastic scatterings are sampled according to a probability $\Gamma_{\text{el}}\Delta t \sim \alpha_s^2 T^3/\mu^2$, where $T$ is the medium temperature and $\mu$ is an infrared regulator given in terms of the Debye mass $m_D$.

The inelastic collisions are initially sampled according to the Gunion-Bertsch (GB) cross-section \cite{Gunion:1981qs}, assuming only a single scattering with the medium. The resulting emitted gluons provides an initial trial (or preformed) gluon which is not yet considered realised, but rather acts as a field coating for the initial projectile parton. We call this the \emph{initial Gunion-Bertsch seed} for the medium-induced radiation. The Gunion-Bertsch energy spectrum applies in the low energy (Bethe-Heitler) regime,
\begin{equation}
	\omega\frac{\mathrm{d}N^{\text{GB}}}{\mathrm{d}\omega} \simeq \frac{\alpha_s C_R}{\pi} \frac{L}{\lambda} \,, \label{eq:gb}
\end{equation}
where $C_R$ is a colour factor ($C_A = 3$ for gluons, $C_F = 4/3$ for quarks), $L$ is the path length of the medium, and $\lambda$ is the Coulomb mean free path $\lambda \sim 1/(\alpha_s T)$ of the medium.

\begin{figure}[ht]
	\centering
	\begin{minipage}{0.49\textwidth}
		\centering
		\begin{tikzpicture}
			\node[anchor=south west,inner sep=0] at (0,0) {\includegraphics[angle=-90,width=0.9\textwidth]{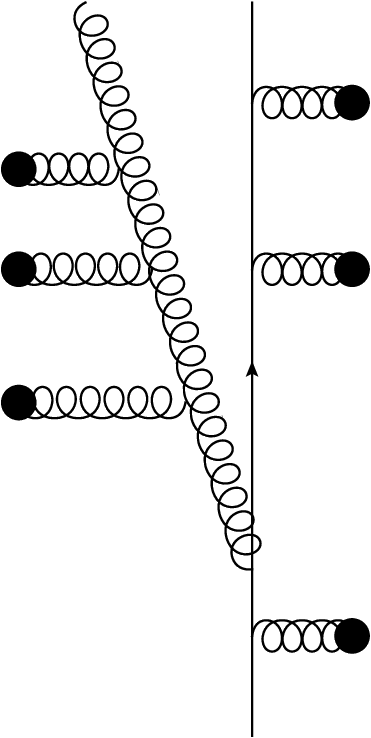}};
			\node at (0,1.4) {$E$};
			\node at (6.5,3.0) {$\omega$};
			\node at (6.5,2.0) {$k_T$};
			\node at (4,4.0) {$N_s$};
			\node[rotate=180] at (4.07,3.6) {{\makebox[15ex]{\upbracefill}}};
			\node[circle,fill,inner sep=0.8pt] at (3.43,2.6) {};
			\node[circle,fill,inner sep=0.8pt] at (3.63,2.6) {};
			\node[circle,fill,inner sep=0.8pt] at (3.83,2.6) {};
		\end{tikzpicture}
		\caption{Illustration of the LPM effect for coherent gluon radiation and some of the relevant variables.}
		\label{fig:lpm}
	\end{minipage}\hfill
	\begin{minipage}{0.49\textwidth}
		\begin{tikzpicture}[domain=0:3]
		\draw[->] (0,0) -- (6,0) node[right] {$\omega$};
		\draw[->] (0,0) -- (0,3) node[above] {$\omega\frac{\mathrm{d}N}{\mathrm{d}\omega}$};
		\draw[color=blue] (0,2.2) node[left,color=black] {$L/\lambda$} -- node[above] {GB} (1.5,2.2);
		\draw[color=green!50!black!80] (1.5,2.2) -- node[below=0.4cm] {BDMPS-Z} (4,1.2);
		\draw[color=orange] (4,1.2) -- node[right=0.5cm] {GLV} (5,0.4);
		\draw[dashed] (1.5,2.2) -- (1.5,0) node[below] {$\lambda m_D^2$};
		\draw[dashed] (4,1) -- (4,0) node[below] {$\omega_c$};
		\draw[dashed] (1.5,2.2) -- (3.5,2.2);
		\draw[->] (3.4,2.0) -- node[right] {$\frac{1}{\sqrt{\omega}}$ suppression} (3.4,1.6);
		\end{tikzpicture}
		\caption{Schematic diagram of the energy radiation spectrum and the different theoretical regimes.}
		\label{fig:energyspectrum}
	\end{minipage}
\end{figure}

However, since the formation of the radiated gluons is a quantum mechanical process, it takes some time to be realised. This is the \emph{formation time} $t_f \sim \sqrt{\omega}$, which is energy dependent. It will therefore be necessary to take into account coherency effects and multiple scatterings with medium. This gives rise to the QCD version of the Landau-Pomeranchuk-Migdal (LPM) effect \cite{Landau:1953um,Migdal:1956tc}. The BDMPS-Z calculation \cite{Baier:1996kr,Baier:1996sk,Zakharov:1996fv,Zakharov:1997uu} of the coherent gluon radiation from a projectile parton, gives rise to the energy spectrum,
\begin{equation}
	\omega\frac{\mathrm{d}N^{\text{BDMPS-Z}}}{\mathrm{d}\omega} \simeq \frac{2\alpha_s C_R}{\pi} \displaystyle \sqrt{\frac{\omega_c}{2\omega}} \,, \hspace{3mm} \text{ for } \hspace{3mm} \omega \ll \omega_c \,, \label{eq:bdmpsz}
\end{equation}
with $\omega_c = \frac{1}{2}\hat q L^2$. Eq.~(\ref{eq:bdmpsz}) shows the characteristic $1/\sqrt{\omega}$ suppression due to the LPM effect. A schematic diagram of the energy spectrum in different regimes is shown in Figure~\ref{fig:energyspectrum}.

In order to achieve coherency, we employ a similar method to JEWEL \cite{Zapp:2011ya}, where for each time-step, the trial gluons can undergo elastic scatterings according to $\Gamma_{\text{el}}\Delta t$ as well as phase accumulation, $\phi \to \phi + \Delta \phi$. Once the phase of the trial gluon has reached a critical threshold value $\phi_c$, the trial gluon is realised with probability $1/N_s$ where $N_s \sim t_f / \lambda$ is the number of elastic scatterings that the trial gluon has undergone during its formation time $t_f$. This will reproduce the BDMPS-Z expectation in eq.~(\ref{eq:bdmpsz}).  


\section{Reproduction of the GB and BDMPS-Z result}

In order to verify our approach, we have considered a static brick medium with constant temperature $T = 400$ MeV, fixed strong coupling $\alpha_s = 0.4$, and path length $L = 4$ fm. The initial jet partons are from a mono-energetic quark gun with $E_{\text{ini}} = 100$ GeV.

\begin{figure}[ht]
	\centering
	\begin{minipage}{0.49\textwidth}
		\centering
		\includegraphics[width=1.02\textwidth]{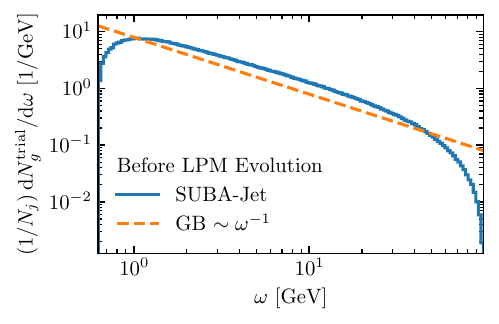}
		\caption{Radiation energy spectrum of trial gluons from SUBA-Jet, right after the emission of a virtual gluon, i.e.\ before the LPM evolution.}
		\label{fig:omegagb}
	\end{minipage}\hfill
	\begin{minipage}{0.49\textwidth}
		\centering
		\includegraphics[width=1.02\textwidth]{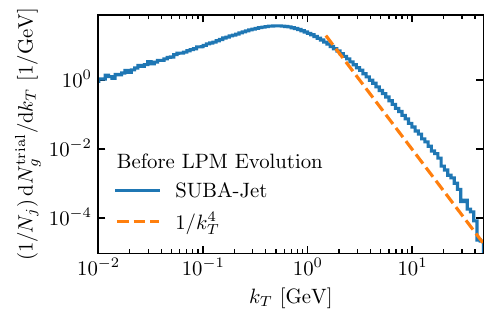}
		\caption{Trial gluon $k_T$ spectrum from SUBA-Jet, right after the emission of a virtual gluon, i.e.\ before the LPM evolution.}
		\label{fig:ktgb}
	\end{minipage}
\end{figure}

In figure~\ref{fig:omegagb}, we present the energy spectrum of the radiated gluons, $\mathrm{d}N/\mathrm{d}\omega$, before the LPM evolution, i.e.\ the trial gluons right after their formation by the initial GB seed. SUBA-Jet reproduces the expected $\omega^{-1}$ behaviour from GB, verifying the initial GB seed procedure.

In figure~\ref{fig:ktgb}, we present the distribution in transverse momentum, $\mathrm{d}N/\mathrm{d}k_T$, of the trial gluons before LPM evolution. For large $k_T$ values, SUBA-Jet nicely reproduces the expected $k_T^{-4}$ behaviour from the GB result.

\begin{figure}[ht]
	\centering
	\begin{minipage}{0.49\textwidth}
		\centering
		\includegraphics[width=1.02\textwidth]{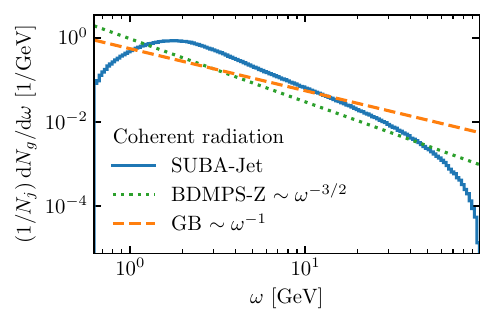}
		\caption{Radiation energy spectrum of gluons from SUBA-Jet, in comparison to the BDMPS-Z results.}
		\label{fig:omegabdmpsz}
	\end{minipage}\hfill
	\begin{minipage}{0.49\textwidth}
		\centering
		\includegraphics[width=1.02\textwidth]{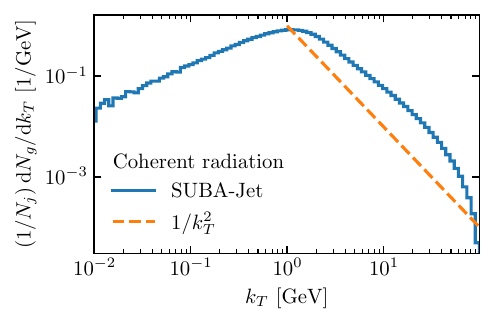}
		\caption{Gluon $k_T$ spectrum from SUBA-Jet, in comparison to the BDMPS-Z results.}
		\label{fig:ktbdmpsz}
	\end{minipage}
\end{figure}

Figures~\ref{fig:omegabdmpsz}~and~\ref{fig:ktbdmpsz} presents the distributions after the LPM evolution, i.e.\ with full coherency. For intermediate energies, SUBA-Jet shows very nice agreement with the BDMPS-Z expectation of $\omega^{-3/2}$ as well as reproduces the expected $k_T^{-2}$ behaviour for large $k_T$. The average number of emitted gluons per jet is around $3$--$4$.

\section{Conclusions and Outlook}

We have presented SUBA-Jet, a new model and Monte Carlo implementation for jet quenching. SUBA-Jet nicely reproduces the BDMPS-Z result. In ref.~\cite{paper} the various model parameters will be varied, and the underlying assumptions of the BDMPS-Z result ($k^{+}$ conservation, infinite scattering centres, etc.) will be relaxed to produce more realistic results.

Future studies will include an interface to vHLLE~\cite{Karpenko:2013wva} to simulate a realistic expanding medium with hydrodynamic evolution, as well as hadronisation and jet finding procedures to produce $p_T$ and $R_{AA}$ distributions. Furthermore, the effect of the jet on the medium will be studied in detail.

The goal is to implement the jet energy loss algorithm from SUBA-Jet in the new EPOS4~\cite{Werner:2023zvo}, so that the initial state, hydrodynamic evolution, and hadronisation will be provided by EPOS4. This is in order to get realistic results which can be directly compared to LHC and RHIC data.


\bibliographystyle{JHEP}
\bibliography{jetrefs}


\end{document}